# Room temperature cathodoluminescence quenching of $Er^{3+}$ in AlNOEr

V. Brien[*1] P. R. Edwards[2], P. Boulet[1], K. P. O'Donnell[2]

1. Institut Jean Lamour, UMR 7198, CNRS, Université de Lorraine, Campus Artem, 2 allée André Guinier, BP 50840, 54011 NANCY, France

2. SUPA Department of Physics, University of Strathclyde, 107 Rottenrow, Glasgow, G4 0NG, Scotland, United Kingdom.

## Abstract

This paper reports a cathodoluminescence (CL) spectroscopic study of nanogranular $AlNOEr_x$ samples with erbium content, $x$, in the range 0.5–3.6 atomic %. A wide range of erbium concentration was studied with the aim of understanding the concentration quenching of CL. The composition of thin films, deposited by radiofrequency reactive magnetron sputtering, was accurately determined by Energy Dispersive X-ray Spectroscopy (EDS). CL emission was investigated in the extended visible spectral range from 350 nm to 850 nm. The critical concentration of luminescent activator $Er^{3+}$ above which CL quenching occurs is 1 %; the corresponding critical distance between $Er^{3+}$ ions in $AlNOEr_x$ is about 1.0 nm. The quenching mechanism is discussed. We discount an exchange-mediated interaction in favour of a multipole-multipole phonon-assisted interaction.

**PACS Code:** 78.60.Kk / 81.15.Cd / 81.05.Ea/ 77.55.hd

**Keywords** : Aluminium nitride; cathodoluminescence; rare earth, quenching; R. F. sputtering

---

[*] Corresponding author. Tel : +33 (0)383684928. E-mail address : valerie.brien@univ-lorraine.fr

1. Introduction

The study of the luminescent properties of rare-earth (RE) doped materials is strongly motivated by their possible technological applications in opto-electronics, colour displays and planar waveguides (information transmission domain)[1]. Erbium (Er)-doped silica is already widely used in optical fibres because its light emission at 1.54µm, due to the transition from the $^4I_{13/2}$ level to the ground-state $^4I_{15/2}$ of the $Er^{3+}$ ion coincides with the low-loss optical window of the fibre. In consequence, the behaviour of the RE has not been studied so extensively in the visible domain.

The technological applications of solid-state light sources require the highest possible optical emission intensities. It is therefore important to optimize the molar fraction of the RE incorporated in the host. Due to its very attractive physical properties (chemical inertia, high melting temperature, a thermal conductivity close to that of copper), aluminium nitride presents itself as the most suitable solid host.

AlNOEr films may be prepared using a sputtering deposition technique well established in industry. The optimization of the dopant fraction is then to be done. In this context, an understanding of the physical reasons governing the optical efficiency, and more specifically its strong decrease when the concentration of the RE increases— called concentration quenching— is a key issue in making doped nitrides attractive for technological applications.

We will first describe the experimental details, and then present the cathodoluminescence (CL) spectra recorded from AlNOEr samples with 0.5 to 3.6 atomic % of erbium; CL spectra will be exploited in order to discuss the luminescence concentration quenching characteristics and mechanisms.

2. Elaboration and structural characterizations of samples

The AlNOEr samples are 500 nm thick films deposited on silicon substrates by R.F. reactive magnetron sputtering. Complete information on the synthesis process and the growth modes of the samples can be found in [2–4]. The resulting samples are poly-granular, exhibiting a columnar morphology with an average grain width of 20–30 nm as shown on the Conventional Transmission Electron Microscopy (CTEM) image, Fig. 1. The variation of the average erbium content from one grown layer to the next was obtained by using an aluminium target on which erbium pellets were placed. In successive growth runs, the depletion of the dopant ensured the samples became progressively poorer in the RE. The morphology of the samples (shape and size of columns) was checked and found to be stable, i.e. independent of Er content. Five samples were prepared with an oxygen content limited to 10 at% maximum by adjusting the base pressure of the reactor; this level is expected to be similar in all samples. After deposition all samples were annealed under $N_2$ (1 bar) at 950°C for 90 min to activate the luminescent centres.

The chemical composition of the samples was assessed primarily by EDS. The analysis was performed on thin sections by means of a Princeton Gamma-Tech (PGT) spectrometer mounted on a Philips CM20 transmission electron microscope equipped with an ultra-thin window X-ray detector. The analyses were carried out in nanoprobe mode with a probe diameter of 10 nm. Such analysis is known to give results with a 10 % precision of the measured value. All films were found to be spatially homogeneous. The compositions of the five samples used in the study are compiled in Table 1. The RBS analyses were performed in the MeV region producing a depth resolution of 5 nm. The discrepancy of the analysis is 5 % of the measured values and was performed to improve the precision on the dosing of Er (the spectroscopy is not so precise on light elements).

3. Cathodoluminescence measurements

Room temperature CL spectra were acquired in a modified Cameca SX100 electron microprobe analyser [5] using a 10 keV, 10 nA electron beam, defocused to a spot diameter of 5 μm in order to maintain a low carrier injection density. An electron energy loss profile was estimated using the Everhart-Hoff approximation [6] (as published for AlN in [10]) and showed that the maximum electron penetration depth is about 500 nm, matching the current AlN film thickness. The emission was dispersed using a ¼-m focal length spectrograph with a 25 μm entrance slit and a 400 lines/mm grating blazed at 500 nm. This allowed detection over the entire visible region using a cooled CCD camera with 25 μm pixel pitch.

Room temperature CL spectra of columnar AlNOEr are presented in Fig. 2. Fig. 2a covers the entire spectral range. The spectra using a logarithmic intensity scale, presented in Fig. 2b and 2c, allow indexing of the emission peaks. Several groups of well resolved sharp peaks belong to the $Er^{3+}$ intra-4f transitions from different excited states to various other levels. We use here the Russell-Saunders notation to refer to the transitions between f states ($^{2S+1}L_J$, where S stands for spin, L for orbital, J for angular momentum quantum numbers). Some of the peaks were identified with the help of fits made by Gruber [7] and are from the $^2P_{3/2}$ level to various other levels. These correspond to the sharper lines seen in our spectra. Most of the other lines involve transitions to the $^4I_{15/2}$ ground state. These values were estimated from a table in Carnall *et al.* [8]. Fig. 3 presents the energy diagram of the Er ion to show the indexing. The only line which does not fall into one of the above categories is that at 707 nm, which could possibly involve a second order diffraction.

4. Discussion

The dominant transitions of our spectra are located at 410 nm, 480 nm and 530–570 nm. Previous CL work recorded at 300 K on AlN:Er nanoparticles deposited on a Si wafer [9] found the 670 nm ($^4F_{9/2} \rightarrow {}^4I_{15/2}$) line to be dominant and the 560 nm peak : $^4S_{3/2} \rightarrow {}^4I_{15/2}$ nearly non-existent, as is commonly observed for photoluminescence (PL) spectra regardless of the sample fabrication process. The intensity ratio of the 530–590 nm peaks to the 650–700 nm peaks (to follow spectral divisions performed in a previous paper [9]) is inverted in the present results; this inversion probably reflects the different morphology of our samples, viz. columnar vs. nanoparticulate.

The erbium concentration dependence of the integrated line intensities is plotted in Fig. 4. Integration was performed either on the whole studied range, or by selecting regions including some of the most emissive peaks, or by selecting also peaks composed of a unique transition. The regions 400 nm – 434 nm, 470 nm – 490 nm and 650 nm – 700 nm correspond to single transitions $^4I_{13/2} - {}^2P_{3/2}$, $^4I_{11/2} - {}^2P_{3/2}$ and $^4F_{9/2} - {}^4I_{15/2}$ respectively. The region (530 nm – 590nm) includes the $^2H_{11/2} - {}^4I_{15/2}$ and $^4S_{3/2} - {}^4I_{15/2}$ and $^4I_{9/2} - {}^2P_{3/2}$ transitions. The optical emission has been normalized to represent optical efficiency *per centre*.

The concentration quenching of RE (RE) doped silica is often attributed to the agglomeration of RE ions in RE-rich nanodomains [1]. Our results and other published works [10] suggest that aluminum nitride does not behave in a similar way to silica. Preliminary results on quenched samples obtained by combining CTEM investigations and X-ray diffraction (XRD) showed no evidence of erbium-rich clusters; in fact, the samples appear to form a solid solution [11]. XRD patterns obtained on the 1% and 3.6% samples are reproduced here (cf. Fig. 5) to facilitate understanding. These confirm the absence of extra phases other than the wurtzite. The experimental investigation was recently complemented by Density Functional Theory (DFT) calculations that exploit the first law of thermodynamics [12] to minimize the chemical potential of model phases. A random distribution of Er atoms in AlN wurtzite, i.e. a solid solution, was shown to be energetically more favorable than the formation of Er nanoclusters, with a total absence of extra phases in the thermodynamical balance up to 12.5 atomic % of erbium. It is then very difficult to propose the existence erbium rich precipitates in the AlN:Er system to justify the observed luminescence concentration quenching, and one has to look for other reasons. A confirmation of the absence of Er-Er bonds in these samples, obtained by using EXAFS on the Er edge, can be found in Ref [13].

In the mid-20$^{th}$ century, the concentration quenching of diluted systems was extensively studied and modelled [14–16] and the energy transfer theories of Förster, Dexter and Schulman are still commonly appealed to today to interpret the physical phenomena, notably by Benz *et al.* who worked on AlN:Er; AlN:Th, AlN:Tb [10,17,18] and by Abhilash Kumar *et al.* who worked on Eu doped oxides [19]. In these models, concentration quenching results from non-radiative energy transfer between luminescence centres. When the $Er^{3+}$ concentration increases, the probability of energy transfer from one centre to another increases as the distance between them decreases. The mechanism responsible for the transfer can be of quantum mechanical type [15,16] and/or electrostatic via multipole–multipole interactions [14]. The value of the critical distance below which the quenching is experimentally observed makes a choice of mechanism possible.

The estimation of average distance *r* between two Er atoms as a function of RE concentration *x* has been carried out on the current data. The calculation was based on the hypothesis of AlNO:Er as a solid solution, with a total dilution of Er. It can be performed in two ways. The value of the specific gravity of AlN (taken at 3.28) can be expressed versus the molar mass of AlNOEr$_x$ that can be computed from the concentration measured for each sample and a spherical volume $4/3\pi$ $(r/2)^3$ occupied by one atom. It could either be computed by using crystallo-chemical considerations taking the number of atoms in a wurtzite lattice and its theoretical volume, by using the cell parameters obtained by XRD in this work .

As can be seen in Fig. 4 the critical concentration of activator is 1%, which leads to a value for critical CL distance CL $r_{critCL}$ of 1.0 nm (6 atomic % of erbium gives 0.56 nm). The exchange interaction mechanism (called the superexchange interaction by Mironov [20]) is a quantum mechanical short-range exchange interaction, becoming ineffective when the distance between phosphor species is larger than 0.5 nm [19,21]. The CL concentration quenching of AlNOEr starts at a greater distance: the exchange interaction mechanism can therefore be discarded over the 0.5 – 1.0 nm range. An electrostatic multipole–multipole interaction [14] can then be invoked to account for the concentration quenching. In that case, the power of this interaction has a direct effect on the decrease of the luminescence intensity and its type can be deduced from the decay of the luminescence intensity. Indeed the emission intensity $I$ (integration of a given peak or identified transition) should follow equation (9) of [19] :

$$\frac{I}{x} = \frac{K}{1+\beta x^{\frac{s}{3}}} \quad (1)$$

$K$ and $\beta$ are constants at the same excitation condition for a given material. Depending on the multipole interaction, the mode of ernergy transfer $s$ is equal to 6, 8, or 10. Some simplification of this expression can be carried out by considering the fact that non-radiative interactions are statistically more probable than radiative ones. After simplifying, and noting that $\alpha=\ln K-\ln\beta$, the log of this gives

$$\ln\left(\frac{I}{x}\right) = \alpha - \frac{s}{3}\ln(x) \quad (2)$$

Logarithmic plots of the visible cathodoluminescence of AlNOEr$_x$ as a function of ln(x) are given in Fig. 6 in order to find *s*. The exploitation of data was done by integrating on different sectors. The slope and the determination coefficient *R* of the basic regression analysis performed on the 4 distribution of points are compiled in Table 2. Except from the region 470 nm - 490 nm which is one of the most intense emissions, *s* is either 6, 8 or 10 (with a satisfying coefficient *R* above 80%) leading to a dipole-dipole interaction for the $^4F_{9/2} - {}^4I_{15/2}$ transition (650 nm - 700 nm sector), to a dipole-quadrupole interaction for the $^2H_{11/2} - {}^4I_{15/2}$ and $^4S_{3/2} - {}^4I_{15/2}$ and $^4I_{9/2} - {}^2P_{3/2}$ transitions (530 nm - 590 nm sector) or to a quadrupole-quadrupole interaction for the $^4I_{13/2} - {}^2P_{3/2}$ (400 nm - 434 nm sector). The obtained coefficient of 14 for the 470 nm - 490 nm corresponding to $^4I_{11/2} - {}^2P_{3/2}$ transitions could correspond to interactions between dipoles of more complex symmetries (sextupole/octopole), but the result is open for discussion.

5. Conclusion

The CL emission of $Er^{3+}$ in polycrystalline nanocolumnar AlNOEr films was studied in the visible domain with particular attention to the concentration quenching behaviour. Quenching of cathodoluminescence emission occurs for $Er^{3+}$ content above 1 % in AlNOEr$_x$. The critical energy transfer distance is found to be 1.0 nm. Based on the experimental results and theoretical calculation, the electrostatic multipole interaction is identified as the major mechanism in concentration quenching, validating the application of Förster's theory.

Acknowledgments

We wish to thank S.S. Hussain, P. Pigeat IJL who helped with the preparation of the samples. The chemical analyses were performed by J. Ghanbaja at the "Service Commun de Microscopie" of the Lorraine University (France). This research did not receive any specific grant from funding agencies in the public, commercial, or not-for-profit sectors.

Table 1. Chemical composition of AlNOEr samples studied here obtained by EDS (precision: +/- 10 % of measured values) or RBS techniques (precision: ± 5 % of measured values). Wurtzite lattice parameters were extracted from 4 circles diffractometry XRD data published in [11]. * EDS calibration was done by using discrete compounds and checked by RBS subsequent analysis. ** Composition here was interpolated by taking the average between samples elaborated before and after, as the elaboration is processing sample depositions one after the other.

| Er content (at %) | EDS* | RBS | a (nm) ± 0.002 | c (nm) ± 0.002 |
|---|---|---|---|---|
| 0.5 | $Al_{41.1} N_{50.1} O_{8.2} Er_{0.6}$ | $Al_{42.5} N_{52} O_5 Er_{0.5}$ | 0.312 | 0.499 |
| 1 | $Al_{41} N_{49.9} O_{8.0} Er_{1.0}$ | Average on the whole thickness: | 0.314 | 0.50 |
| 1.4 | - | **$Al_{43} N_{49} O_6 Er_{1.4}$** | 0.314 | 0.50 |
|  |  | $Al_{43} N_{49} O_7 Er_1$ on the top 3/5 of the film |  |  |
|  |  | $Al_{43} N_{49} O_6 Er_2$ on the last 2/5 underneath |  |  |
| 2.8 | $Al_{41.4} N_{49.8} O_6 Er_{2.8}$** |  | 0.315 | 0.502 |
| 3.6 | $Al_{40.2} N_{51.1} O_{5.1} Er_{3.6}$ |  | 0.316 | 0.503 |

Table 2. Values of slopes obtained by linear regression from plots of integrated CL data of single transitions visible in Fig. 6. Deducted s coefficient ($s = 3 \cdot slope$).

| Sectors | 400 nm - 434 nm | 470 nm - 490 nm | 530 nm - 590 nm | 650 nm - 700 nm |
|---|---|---|---|---|
| slope | 3.35 | 4.58 | 2.80 | 2.09 |
| s | 10.1 | 13.7 | 8.4 | 6.3 |
| Closest integer | 10 | 14 | 8 | 6 |
| $R^2$ | 0.8146 | 0.9172 | 0.8396 | 0.9259 |

Fig. 1: Typical TEM cross-section recorded on the AlNO:Er$_{x(x=0.5–3.6at.\%)}$, polycrystalline sputtered films of columnar morphology. Dark field image.

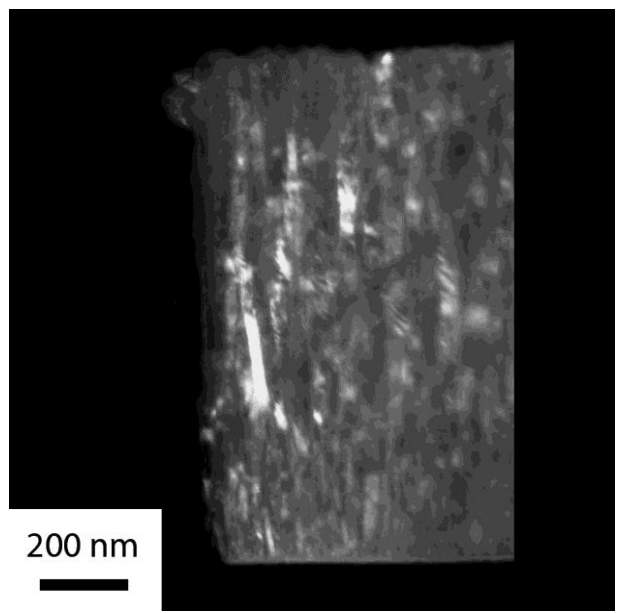

Fig. 2: Room temperature cathodoluminescence in the visible range of AlNOEr films synthesized by sputtering as a function of the content of Erbium $x$. CL was performed at 300K using electron beam energy of 10 keV, current of 10 nA defocusing to a spot diameter of 5 µm. a/ Entire spectra. $x$ is increasing from bottom spectrum to top one; b, c/ Logarithmic plots with indexation of emission peaks.

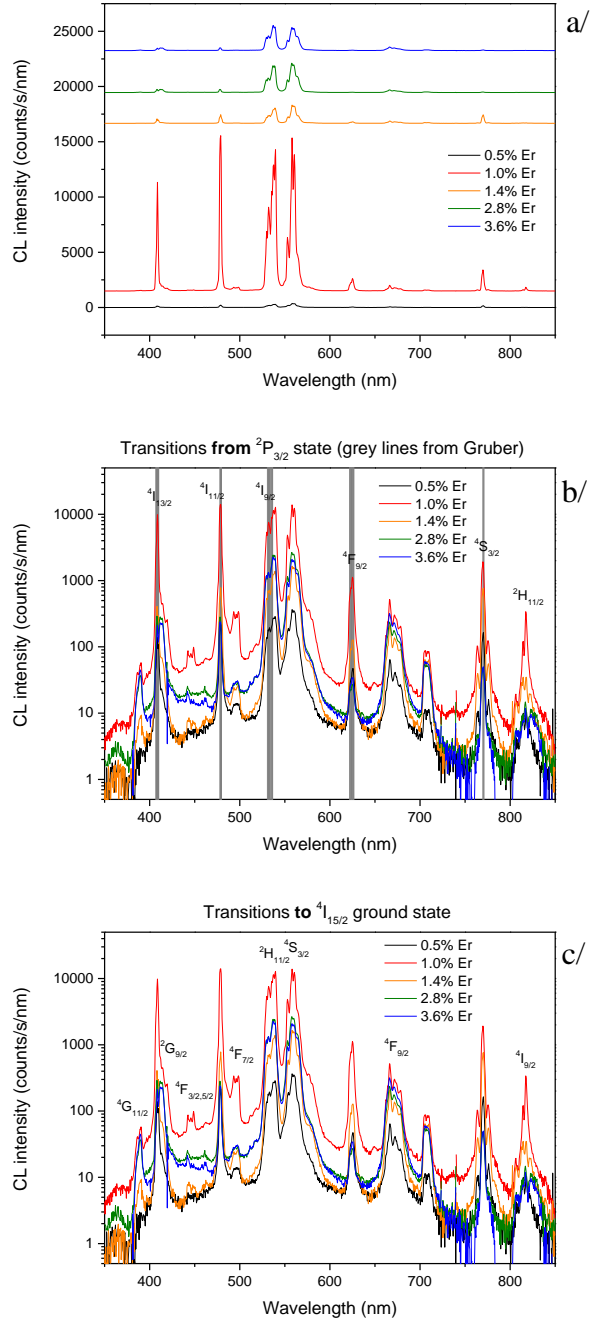

Fig. 3: Energy diagram of the Er ion after *Carnall et al.* [24] to visualize the corresponding transitions.

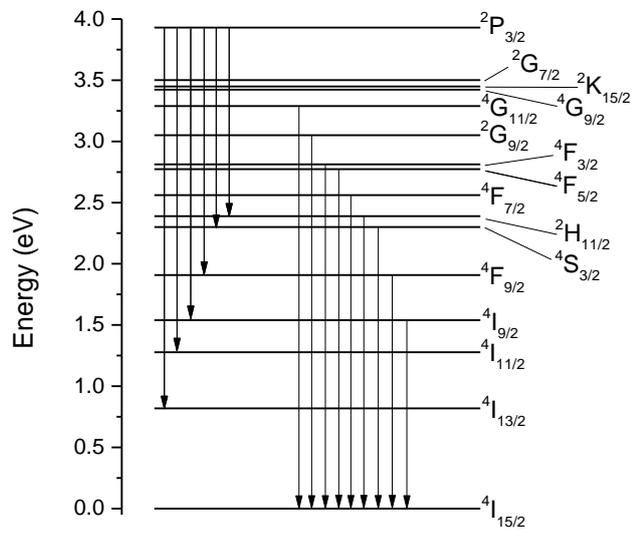

Fig. 4: Optical CL efficiency of emitters (global integration on mentioned sectors) as a function of the $Er^{3+}$ concentration $x$.

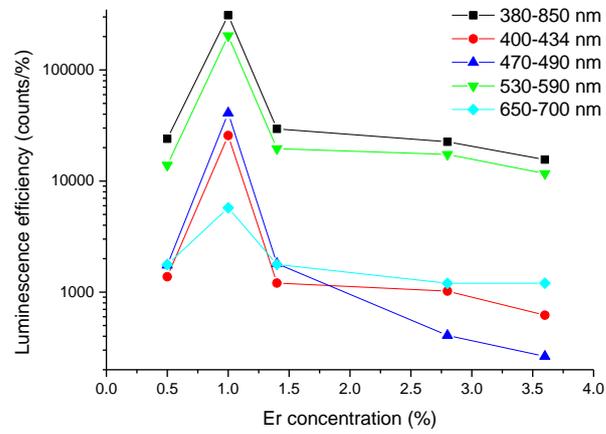

Fig. 5 : Cumulated XRD patterns obtained by four circles diffractometry on $AlNOEr_{x(x=1\ and\ 3.6\%)}$ films. Bottom data is related to the naked silicon substrate.

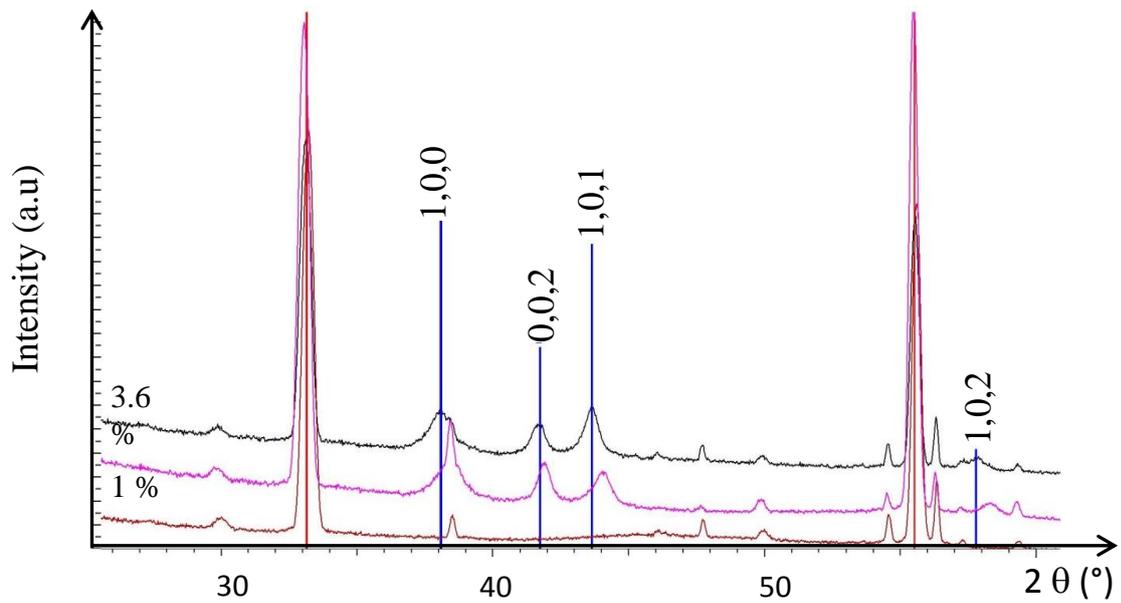

Fig. 6: Logarithmic plots of cathodoluminescent results of integrated data on the same sectors as in spectroscopic records given in Fig 4.

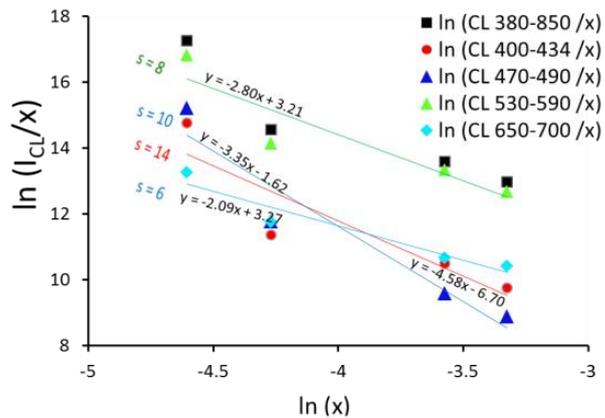